\documentstyle[aps,twocolumn,epsfig]{revtex}
\begin{document}
\title{Superluminal Signals: Causal Loop Paradoxes Revisited}
\author{J.  C.  Garrison, M.  W.  Mitchell , 
and R.  Y.  Chiao \\
University of California\\
Berkeley, CA 94720}
\author{E.  L.  Bolda\\University of Auckland\\ Private Bag 
92019, Auckland, New Zealand}
\maketitle
\begin{abstract}

Recent results demonstrating superluminal group velocities and 
tachyonic dispersion relations reopen the question of superluminal 
signals and causal loop paradoxes.  The sense in which superluminal 
signals are permitted is explained in terms of pulse reshaping, and 
the self-consistent behavior which prevents causal loop paradoxes is 
illustrated by an explicit example.

\end{abstract}

\section{Introduction} \label{intro}

The idea of ``tachyons'', i.e., particles that travel in the vacuum 
faster than light, has been the source of controversy for many years.  
Although special relativity does not strictly outlaw tachyons 
\cite{1167,1168,1103}, the interaction of tachyons with ordinary 
matter does raise difficult questions.  One of these is the 
possibility of violating the familiar relativistic prohibition of 
faster-than-light (superluminal) signals.  A closely related concern 
is that \textit{any} interaction of tachyons with ordinary matter 
would lead to logical inconsistencies through the formation of closed 
causal loops \cite{1096,764,835}.  The participants in the ongoing 
debates are at liberty to hold their various views largely because of 
the complete absence of any experimental data.  This unsatisfactory 
situation persists as far as true tachyons are concerned, but not with 
regard to ``quasitachyons'', i.e., excitations in a material medium 
exhibiting tachyon-like behavior.  Theoretical considerations have 
shown that superluminal and even negative group velocities are 
physically meaningful \cite{403,1164,143,144}, and that excitations 
with tachyonic dispersion relations exist \cite{1159,1191,1192,1156}.  
Superluminal group velocities have been experimentally observed for 
propagation through an absorbing medium \cite{248}, for microwave 
pulses\cite{1160,1193,1162} , and for light transmitted through a 
dielectric mirror \cite{1163,1165} .  There has been a parallel 
theoretical controversy over the possibility of superluminal behavior 
in quantum tunneling of electrons and photons.  Recent experiments 
using photons as the tunneling particles \cite{1163,1189,1170,1190} 
have confirmed Wigner's early prediction that the time required for a 
particle to traverse a tunneling barrier of width $d$ can indeed be 
less than $d/c$.

The existence of superluminal group velocities and quasitachyons 
raises questions of the same general kind as those sparked by the 
previous speculations about tachyons.  Is there any conflict with 
special relativity?  Can these phenomena be used to send superluminal 
signals?  What mechanism prevents logical contradictions through the 
formation of closed causal loops?  In order to arrive at reasonably 
sharp and concise answers to these questions we will restrict the 
following discussion primarily to classical phenomena.  The answer to 
the first question is straightforward.  In all cases considered so 
far, the propagation of excitations in a medium is described by 
theories, e.g., Maxwell's equations, which are consistent with special 
relativity; therefore, the predictions cannot violate special 
relativity.  The remaining questions require somewhat more discussion.  
Superluminal signaling will be examined in Sec.~\ref{super}, in the 
context of choosing an appropriate definition of signal propagation 
speed.  In Sec.~\ref{paradoxes} we investigate the issue of causality 
paradoxes in a somewhat simpler context.  Finally the lessons drawn 
from these considerations will be discussed in Sec.~\ref{discuss}.

\section{Superluminal signals} \label{super}

A common, if loosely worded, statement of an important consequence of 
special relativity is:``No signal can travel faster than light.'' The 
more sweeping statement,``Nothing can travel faster than light.'', is 
contradicted by the familiar example of the spot of light thrown on a 
sufficiently distant screen by a rotating beacon \cite{1173}.  The 
apparent velocity of the spot of light can exceed $c$, but this does 
not contradict special relativity since there is no causal relation 
between successive appearances of the spot.  Any discussion of the 
first statement requires a definition of what is meant by a ``signal'' 
and what is meant by ``signal velocity''.  For our purposes it is 
sufficient to define a signal as the emission of a well defined pulse, 
e.g., of electromagnetic radiation, at one point and the detection of 
the same pulse at another point.

The classic analysis of Sommerfeld and Brillouin \cite{1169} 
identified five different velocities associated with a 
finite-bandwidth pulse of electromagnetic radiation propagating in a 
linear dispersive medium.  We will consider here only the ``front 
velocity'', the velocity of the ``front'', i.e., the boundary 
separating the region in which the field vanishes identically from the 
region in which the field assumes nonzero values, and the ``group 
velocity'', which describes the overall motion of the pulse envelope.  
It is worth noting that the definition of the front does not require a 
jump discontinuity at the leading edge of the pulse.  Discontinuities 
of this kind are convenient idealizations suggested by the difficulty 
of following the behavior of the pulse at very short time scales, but 
they can always be replaced by smooth behavior.  An envelope function 
which is sufficiently smooth at the front, e.g., the function and its 
first derivative vanish there, will have a finite bandwidth in the 
precise sense that the rms dispersion of the frequency, calculated 
from the power spectrum, will be finite.  This definition of bandwidth 
is the one used in the uncertainty relation.  The existence of the 
front has an important consequence which is best described in the 
simple example of one-dimensional propagation.  If the incident wave 
arrives at \(x = \mathrm 0\) at \(t = \mathrm 0\) and the envelope, 
\(u\left({\mathrm x,t}\right)\), satisfies \(u\left({\mathrm 
0,t}\right) = \mathrm 0\) for \(t < \mathrm 0\), then physically 
plausible assumptions for the behavior of the medium guarantee that 
the front propagates precisely at \(c\), the velocity of light 
\textit{in vacuo}\cite{544}.  On the other hand, the group velocity 
can take on any value.  Indeed it has been shown that ``abnormal'' 
(either superluminal or negative) group velocities are required by the 
Kramers-Kronig relation for some range of carrier frequencies away 
from a gain line or within an absorption line\cite{143}.

The principle of relativistic causality states that a source cannot 
cause any effects outside its forward light cone.  Since the front of 
a pulse emitted by the source traces out the light cone, this means 
that no detection can occur before the front arrives at the detector.  
A general signal will be a linear superposition of the elementary 
signals described above.  The simplest model of this general behavior 
is that the pulse envelope \(u\left({\mathrm x,t}\right)\) is 
sectionally analytic in \(t\), i.e., the front is a point of 
nonanalyticity separating two regions in which the pulse envelope has 
different analytic forms.  The values of the pulse envelope on any 
finite segment in the interior of a domain of analyticity determine, 
by the uniqueness theorem for analytic functions, the pulse envelope 
up to the next point of nonanalyticity.  It is therefore tempting to 
associate the arrival of new data with the points of nonanalyticity 
\cite{1170} in the pulse envelope.  From this point of view, it is 
reasonable to identify the signal velocity with the front velocity.  A 
happy consequence of this choice is that superluminal signals are 
uniformly forbidden, but this conceptual tidiness is purchased at a 
price in terms of experimental realism.  By definition, the field 
vanishes at the front, and for smooth pulses will remain small for 
some time thereafter.  Thus the front itself cannot be observed by a 
detector with finite detection threshold.  Nevertheless, an 
operational definition of the front velocity can be given in terms of 
a limiting procedure in which identical pulses are detected by a 
sequence of detectors, \({D}_{n}\) , with decreasing thresholds 
\({S}_{n}\) .  Let the pulse be initiated at \(t = \mathrm 0\) and 
denote by \({t}_{n}\) the time of first detection by \({D}_{n}\) , 
then the effective signal velocity is \({v}_{n} = d / {t}_{n}\),where 
$d$ is the distance from the source to the detector.  The front 
velocity would then be defined by extrapolating \({v}_{n}\) as 
\({S}_{n}\mathrm \rightarrow \mathit \mathrm 0\).  While physically 
and logically sound, this procedure is scarcely practical.

We now turn from consideration of a series of increasingly sensitive 
detectors to a single detector with threshold close to the expected 
peak strength of the signal, \({S}_{\mathrm 0}\mathrm \ \approx 
\mathit \left|{{u}_{peak}}\right|\) .  We also assume that the pulse 
is not strongly distorted during propagation.  Under these 
circumstances the pulse envelope propagates rigidly with the group 
velocity \({v}_{g}\).  This suggests identifying the signal velocity 
with the group velocity.  This is an attractive choice from the 
experimental point of view, but this benefit also has a price.  First 
note that the peak cannot overtake the front \cite{1169} and that the 
front travels with velocity \(c\).  This prompts the question.  In 
what sense is the signal superluminal even if \({v}_{g} > c\)?  To 
answer this, consider an experiment in which the original signal is 
divided, e.g., by use of a beam splitter.  One copy is sent through 
the vacuum and the other through a medium, and the firing times of 
identical detectors placed at the ends of the two paths, each of 
length d, are then recorded.  The difference between the two times, 
the ``group delay'', is given by

\begin{equation}{t}_{g} = {\frac{d}{{v}_{g}}} - {\frac{d}{c}}~. 
\label{grpdelay}\end{equation}
The group delay is positive for normal media (\(\mathrm 0\mathit < 
{v}_{g} < c \) ), and negative for abnormal media, (\({v}_{g} > c\) or 
\({v}_{g} < 0\)).  It is only in this sense that the abnormally 
propagated signal is superluminal.  With these definitions it is then 
correct to say that special relativity does not prohibit superluminal 
signals.  This is a fairly innocuous complication of the usual 
discussion, but there are more serious problems related to the 
robustness of the definition.  For example, if a more sensitive 
detector were used the measured group delay could be significantly 
smaller than that given by (\ref{grpdelay}).  Indeed as the threshold 
of the detector approaches zero, the group delay would approach zero.  
In other words the signal velocity would approach the front velocity.  
The only simple way to remove this ambiguity would be to identify the 
arrival of the pulse with the arrival of the peak.  This would seem to 
attribute an unwarranted fundamental significance to the peak.

\section{Causal loop paradoxes}\label{paradoxes}

Causal loop paradoxes are usually introduced by considering two 
observers, A and B, each equipped with transmitters and receivers for 
tachyons.  At time \({t}_{A}\), A sends a tachyonic message to B who 
then sends a return message to A timed to arrive at \({t}_{A}^{\mathrm 
'} < {t}_{A}\), where both times are measured in A's restframe.  The 
paradox occurs if the return message activates a mechanism which 
prevents A from sending the original message \cite{1096}.  Our next 
task is to reexamine this issue in the context of the two definitions 
of signal velocity discussed above.  No paradoxical behavior is 
possible if the signal velocity is identified with the front velocity, 
since the signal velocity then equals the velocity of light.  When the 
signal velocity is equated to the group velocity, more discussion is 
needed, since negative group delays are possible.

\begin{center} 
\begin{figure} 
\psfig{width=3in,figure=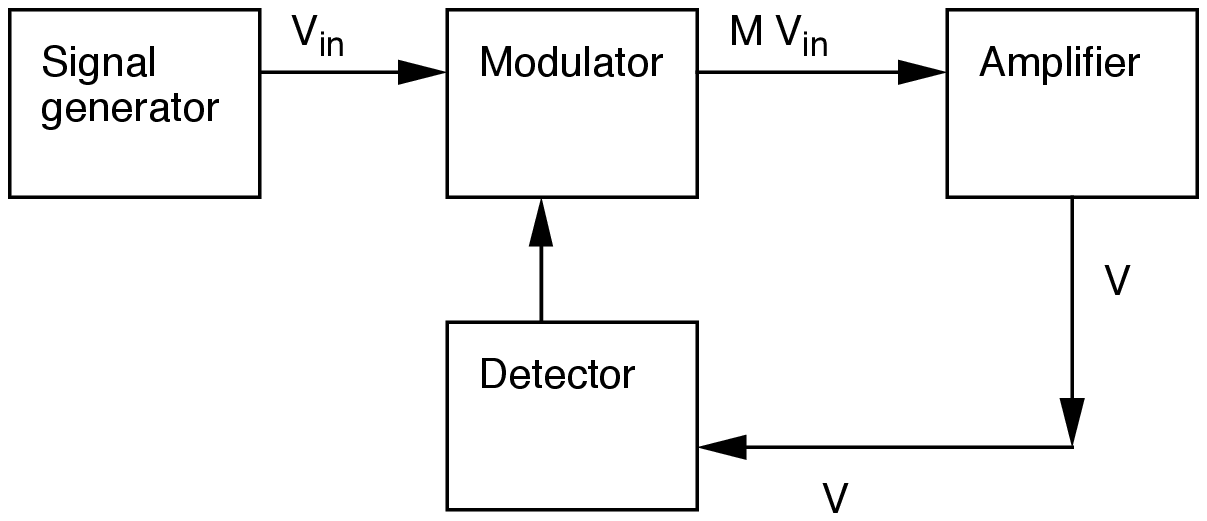}
\caption{Feedback circuit: The modulator multiplies the input signal 
by $M(t)$ when the detector fires.  The amplifier parameters are 
\({\mathrm \omega }_{r} = \mathrm 5 \mathit\mathrm 1 \mathit Hz\) 
,\(\mathrm \gamma \mathit = \mathrm 1 \mathit\mathrm 5 \mathit {s}^{- 
\mathrm 1 \mathit}\) ,\({T}_{\mathrm 0} = \mathrm 2 \mathit.\mathrm 9 
\mathit\mathrm 4 \mathit ms\) }

\label{circuit}
\end{figure} 
\end{center}

The core of the tachyon paradox is the ability to send messages into 
the past.  It is therefore sufficient to devise a situation in which 
messages can be sent to the past at a single point in space 
\cite{764}.  A concrete example can be constructed by using an 
electronic circuit, for which the light transit time across the system 
is negligible compared to all other time constants.  Propagation 
effects are then irrelevant, and the system can be described by a 
function, \(V\left({t}\right)\), which depends only on time.  For 
these systems we can use the principle of elementary causality which 
states that the output signal depends only on past values of the input 
signal.  We will assume that both the input and the output pulse have 
well defined peaks occurring at \({t}_{in}\) and \({t}_{out}\) 
respectively.  The time difference \({t}_{g} = {t}_{out} - {t}_{in}\) 
is called the group delay by analogy to (\ref{grpdelay}).  Analyzing 
the relation between input and output in this way is analogous to 
choosing the group velocity to represent the signal velocity in the 
propagative problem.  We can attempt to create the paradox by 
designing a circuit with negative group delay, i.e., the output peak 
leaves the amplifier \textit{before}the input peak has entered.  A low 
frequency bandpass amplifier with this rather bizarre property has 
been experimentally demonstrated \cite{1158}, and it will be used in 
the following discussion.  To get a message from the future it is 
necessary to construct a feedback loop in which the output of the 
amplifier is used to modulate the input, as shown in 
Fig.~\ref{circuit}.  If the amplifier produces a negative group delay 
this arrangement could apparently be used to turn off the input 
prematurely, e.g., before the peak.

\begin{center} 
\begin{figure} 
\psfig{width=3in,figure=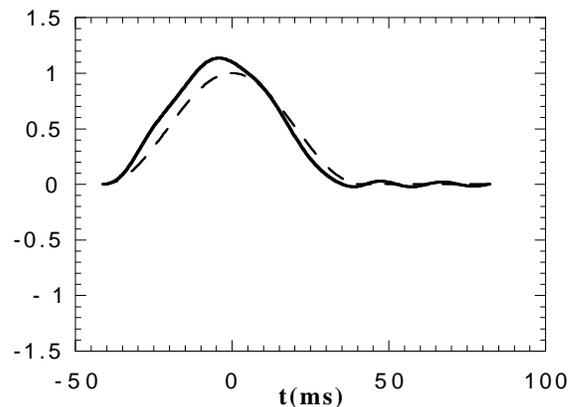} 
\caption{$V_{out}(t)/V_{0}$(solid) and $V_{in}(t)/V_{0}$ (dashed) 
vs.  time.  Input pulse parameters \({\mathrm \omega }_{c} = \mathrm 
0\) , \({T}_{f} = \mathrm 4 \mathit\mathrm 1 \mathit ms\).  This shows 
negative group delay, $t_{g}= - 3.7 ms$, for the parameters of 
Fig\ref{circuit}.}

\label{ngrpdelay}
\end{figure}
\end{center}

The Green's function of the amplifier, i.e., the Fourier transform of 
the frequency-domain transfer function, is

\begin{equation}G\left({t}\right) =\mathrm \ \delta \left({\mathit 
t}\right)\mathit + G\mathrm '\left({\mathit t}\right)~, \end{equation}

\begin{equation}G\mathrm '\left({\mathit t}\right)\mathit \mathrm 
\equiv \mathit {G}_{\mathrm 0} \mathrm \gamma \mathit \mathrm \theta 
\left({\mathit t}\right)\mathit {e}^{-\mathrm \ \gamma \mathit t} 
\left[{\cos \left({{\mathrm \omega }_{r} t}\right) + {\frac{\mathrm 
\gamma }{{\omega }_{\mathit r}}} \sin \left({{\mathrm \omega }_{r} 
t}\right)}\right]~,\label{prop} \end{equation}
where \(\mathrm \theta \left({\mathit t}\right)\) is the step 
function, \(\mathrm \gamma \) and \({\mathrm \omega }_{r}\) are 
respectively the damping rate and resonant frequency of the amplifier, 
and the dimensionless parameter \({G}_{\mathrm 0}\) describes the 
overall amplification \cite{1158}.  The presence of the step function 
in (\ref{prop}) imposes the retarded Green's function and guarantees 
elementary causality.  In the absence of feedback the output signal is

\begin{equation}{V}_{out}\left({t}\right) = {V}_{in}\left({t}\right) + 
\int_{-\mathrm \ \infty }^{t}d\mathrm \tau \mathit G\mathrm 
'\left({\mathit t -\mathrm \ \tau }\right)\mathit 
{V}_{in}\left({\mathrm \tau }\right) \label{vout}~. \end{equation}
A simple example of an input signal which has a continuous first 
derivative everywhere and vanishes outside a finite interval, is 
given by

\begin{equation}{V}_{in}\left({t}\right) = {V}_{\mathrm 0} \mathrm 
\theta \left({{\mathit T}_{\mathit f}\mathit - 
\left|{t}\right|}\right)\mathit \cos \left({{\mathrm \omega }_{c} 
t}\right) {\cos}^{\mathrm 2 \mathit}\left({{\frac{\mathrm \pi \mathit 
t}{\mathrm 2 \mathit {T}_{f}}}}\right)~. \end{equation}
In the interior of the interval \(\left({- {T}_{f} , 
{T}_{f}}\right)\), this signal resembles a Gaussian pulse peaked at 
\(t = \mathrm 0\), and modulated at carrier frequency \({\mathrm 
\omega }_{c}\) .  An example of negative group delay for this input is 
shown in Fig.~\ref{ngrpdelay}.

\begin{center} 
\begin{figure}
\psfig{width=3in,figure=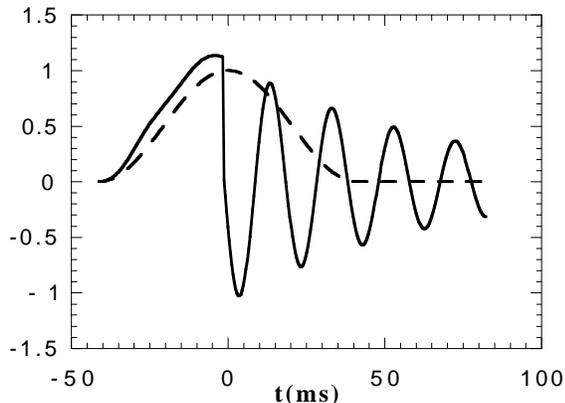} 
\caption{$V(t)/V_{0}$ and $V_{in}(t)/V_{0}$ \textit{vs} time.  
Solution of the feedback equation (\ref{paradox}) with $S_{0}= 1.12 
V_{0} $.  Input signal turned off 1.5 ms before its peak.  Other 
parameters as in Fig.~\ref{circuit} }

\label{fdbksol} 
\end{figure} 
\end{center}

In the feedback circuit the detector, with threshold set to \({S}_{\mathrm 
0}\), triggers the modulator which in turn multiplies the input 
voltage by a factor \(M\left({t}\right)\).  The input to the amplifier 
is then \(M\left({t}\right) {V}_{in}\left({t}\right)\), and the signal 
\(V\left({t}\right)\) satisfies

\begin{eqnarray}V\left({t}\right) &=& M\left({t}\right) 
{V}_{in}\left({t}\right) 
\nonumber \\
&&+ \int_{-\mathrm \ \infty }^{t}d\mathrm \tau 
\mathit G\mathrm '\left({\mathit t -\mathrm \ \tau }\right)\mathit 
M\left({\mathrm \tau }\right) {V}_{in}\left({\mathrm \tau }\right)~. 
\label{feedback} \end{eqnarray}
This seems to open the way for a variant of the time travel paradox in 
which the traveller journeys to the past and kills his grandfather 
before his own father is born.  The analogous situation for the 
feedback circuit would be to employ the output peak to turn off the 
input before it has reached its peak.  Fig.~\ref{ngrpdelay} shows that 
this seems to be possible.  If there is a paradox, the integral 
equation (\ref{feedback}) should fail to have a solution when the 
modulation function is chosen in this way. In an attempt to produce the 
paradoxical situation we choose the modulating function 
\(M\left({t}\right)\) as follows.  For any signal 
\(V\left({t}\right)\) which rises smoothly from zero, define 
\({t}_{\mathrm 1 \mathit}\) and \({t}_{\mathrm 2 \mathit}\) as the 
first two times for which \(\left|{V\left({t}\right)}\right| = 
{S}_{\mathrm 0}\).  The first peak of the amplitude exceeding $S_{0}$ 
is guaranteed to lie between these two times, provided that the value 
of $S_{0}$ is below the absolute maximum value of the feedback 
signal.  The modulating function is then chosen as

\begin{equation}M\left({t}\right) =\mathrm \ \theta \left({{\mathit 
t}_{\mathit \mathrm 2}\mathit - t}\right)~. \end{equation}
Thus the input is unmodulated until the peak of the feedback signal 
has passed and the detector again registers $S_{0}$.  At this time 
the input is set to zero.  The integral equation (\ref{feedback}) for 
the signal is now

\begin{eqnarray}V\left({t}\right) &=& \mathrm \theta \left({{\mathit 
t}_{\mathit \mathrm 2 \mathit}\mathit - t}\right)\mathit 
{V}_{in}\left({t}\right) 
\nonumber \\
&&+ \int_{-\mathrm \ \infty }^{t}d\mathrm \tau 
\mathit G\mathrm '\left({\mathit t -\mathrm \ \tau }\right)\mathit 
\mathrm \theta \left({{\mathit t}_{\mathit \mathrm 2 \mathit}\mathit 
\mathrm \tau }\right)\mathit {V}_{in}\left({\mathrm \tau }\right)~. 
\label{paradox}\end{eqnarray}
For times \(t < {t}_{\mathrm 2}\), the modulation function in both 
terms of (\ref{paradox}) is unity, and comparison with (\ref{vout}) 
shows that \(V\left({t}\right) = {V}_{out}\left({t}\right)\) for \(t < 
{t}_{\mathrm 2}\) .  Thus the time \({t}_{\mathrm 2}\) is determined 
by the simple output function \({V}_{out}\), and the solution to 
(\ref{paradox}) is

\begin{eqnarray}{V}^{}\left({t}\right) &=&\mathrm \ \theta 
\left({{\mathit t}_{\mathit \mathrm 2 \mathit}\mathit - 
t}\right)\mathit {V}_{out}\left({t}\right) 
+ \mathrm \theta 
\left({\mathit t - {t}_{\mathrm 2 \mathit}}\right)\mathit 
\nonumber \\
&&\times \int_{-\mathrm \ \infty }^{t}d\mathrm \tau \mathit G\mathrm 
'\left({\mathit t -\mathrm \tau }\right)\mathit \mathrm \theta 
\left({{\mathit t}_{\mathit \mathrm 2 \mathit}\mathit -\mathrm \tau 
}\right)\mathit {V}_{in}\left({\mathrm \tau }\right)~, \end{eqnarray}
where $t_{2}$ is determined by $V_{out}$.  A computer simulation of 
the solution, using the same parameters as in Fig.~\ref{circuit} , is 
plotted in Fig.~\ref{fdbksol}.  The self-consistent signal follows the 
amplifier output until the detector is triggered.  This occurs after 
the output signal has reached its peak but before the input achieves 
its peak.  The sudden termination of the input then sets off a damped 
oscillation.  The existence of a self-consistent solution shows that 
there is no paradox; i.e., the theory does not suffer from internal 
contradictions.  This feature is shared with previous resolutions of 
apparent paradoxes associated with tachyons \cite{764,835}, or with 
the use of advanced Green's functions in the Wheeler-Feynman radiation 
theory \cite{1157}.

\section{Discussion and conclusions}\label{discuss}

In Sec.~\ref{super} we considered only two candidates for the signal 
velocity.  While they may not be the only possibilities, the front and 
group velocities do seem to have the strongest \textit{a priori} claims.  
Furthermore there is a form of complementarity between them.  The front 
velocity is conceptually simple but operationally complex, and the 
group velocity is conceptually complex but operationally simple.  Each 
alternative has strengths and weaknesses which we now discuss.

Identification of the front velocity as the signal velocity uniformly 
forbids the appearance of any superluminal signals, either in the 
vacuum or in a medium.  This definition is Lorentz invariant, and it 
automatically excludes the possibility of any causal loop paradoxes.  
For a given distance $d$ between source and detector, the predicted 
arrival time $d/c$ represents the earliest possible time for 
detection.  This interpretation is related to the most serious 
drawback of the definition, namely a detector with finite threshold cannot 
respond until some time $t>d/c$.  Thus the arrival time $d/c$ can only 
be approximated by a limiting procedure such as that discussed in 
Sec.~\ref{super}.  This objection is not fatal, since definitions of 
fundamental notions in terms of a limiting procedure are common, e.g., 
the definition of an electric field as the ratio of the force on a 
test charge to the charge as the charge approaches zero.  An 
additional drawback is that the front-velocity definition mandates 
that all signals travel exactly at $c$, whether in the vacuum or in a 
medium. In particular this means that signals travel exactly at $c$ in normal 
dielectrics, not slower than $c$.  This is not in accord with our usual usage and intuitions.

Identification of the group velocity with the signal velocity has the 
advantage of easy experimental realization, but there are also 
disadvantages.  For example, one can imagine signals transmitted by 
pulses lacking a well defined group velocity, e.g., in the 
presence of strong group velocity dispersion.  Furthermore, this 
definition actually requires the existence of superluminal signals, in 
the sense of negative group delays.  In view of this, it is natural to 
wonder how superluminal signals can be consistent with special 
relativity.  To begin, recall that special relativity is based on two 
postulates: (A) The laws of physics have the same form in all inertial 
frames.  (B) The velocity of light \textit{in vacuo} is independent of 
the velocity of the source.  The first postulate is already present in 
Newtonian mechanics, so it is the second that leads to 
characteristically relativistic phenomena.  Neither postulate says 
anything directly about the propagation of excitations in a material 
medium.  The implications of special relativity for this question can 
only be found by using a theory of the medium, e.g., the macroscopic 
form of Maxwell's equations, which is consistent with special 
relativity.  In all such theories the response of the medium to the 
incident wave is described by retarded propagators, in accordance with 
both relativistic and elementary causality.  With this in mind, 
superluminal propagation of electromagnetic fields can be understood 
as reshaping of the pulse envelope by interaction with the medium 
\cite{144}.  For propagation in a linear medium, it has long been 
known that the peak of the pulse can never overtake the 
front\cite{1169}.  This conclusion holds for nonlinear media as well, 
e.g., superluminal propagation in a laser amplifier \cite{534}.  In 
all cases the pulse shape will become increasingly distorted as it 
asymptotically attempts to overtake the front.

The would-be paradoxical feedback circuit analyzed in 
Sec.~\ref{paradoxes} at first appears to present a puzzle.  With the 
choice of parameters in Fig.~\ref{fdbksol}, the peak of the feedback 
signal is used to turn off the input signal before it achieves its 
peak.  This would seem to satisfy the requirements of the paradox, but 
the feedback problem does have a self-consistent solution.  The 
apparent difficulty here stems from the natural assumption that the 
output peak is causally related to the input peak.  This assumption 
has been criticized previously \cite{1173}, and recent experimental 
results \cite{1158} , as well as the simulation results shown in 
Fig.~\ref{fdbksol}, show it to be false.  In order for event A to be 
the cause of event B, it must be that preventing A also prevents B.  
Both experiment and theory show that preventing the peak in the input 
does not prevent the peak in the output, therefore the peaks are not 
causally related.  The peak in the output is however causally related 
to earlier parts of the input, since cutting off the input 
sufficiently early will prevent the output peak from appearing 
\cite{1187}.  This shows that the analytic continuation of an initial 
part of the smooth pulse, discussed in Sec.~\ref{super}, is not just a 
theoretical artifact; the experimental apparatus actually performs the 
necessary extrapolation.  However the apparatus cannot send a signal 
to any time prior to the initiation of the input signal.  In other 
words, the output signal vanishes identically for \(t < - {T}_{f}\); 
this is guaranteed by the use of retarded propagators.

The discussion so far has been carried out at the classical level, but 
there are general arguments suggesting that there will be no surprises 
at the quantum level.  The relevant setting here is quantum field 
theory.  In the Heisenberg picture the operator field equations have 
the same form as the classical field equations, so it is plausible 
that the solutions will be described by the same propagators.  In 
particular, the electromagnetic field operators arising from an 
electric current localized in a small space-time region will be 
related to the source by the standard retarded propagator which 
vanishes outside the light cone.  Indeed the solution to the point 
source problem involves only the retarded propagator even for models 
with tachyonic dispersion relations \cite{10}.  Explicit calculations 
for one such model display the same pulse reshaping features as the 
classical case \cite{1176}.  A rigorous, general argument has been 
given by Eberhard and Ross \cite{1154}, who show that if classical 
influences satisfy relativistic causality, then no signals outside the 
forward light cone will be observed in a fully quantal calculation.  
The essential point for their argument is the postulate that operators 
localized in space-like separated regions commute.  This is used to 
show that actions performed in one region cannot change the 
probability distributions for measurements in a space-like separated 
region.

The first conclusion to be drawn from this discussion is that there is 
no completely compelling argument that would allow a choice between 
the proposed definitions of signal velocity.  The front-velocity 
definition eliminates all superluminal signals and causality problems 
at a single stroke, but at the expense of an indirect operational 
definition.  The group-velocity definition is operationally simple, 
but it provides a well defined sense in which superluminal signalling 
is allowed by relativity and by elementary causality, namely in those 
media allowing negative group delay.  The description of negative 
group delay as superluminal propagation is, to some extent, a question 
of language.  The values of the group delay, whether positive or 
negative, come from pulse reshaping effects.  Thus one could speak of 
``group advance'' for abnormal propagation and ``group 
retardation'' for normal propagation.  The choice between 
``superluminal signal'' and ``group advance'' is a matter of taste, 
but it should be kept in mind that the group delay is a measurable 
quantity and that negative values have been observed.  Another point 
to consider is that before the work of Garrett and McCumber\cite{403} 
the possibility of negative group delays would have been rejected as 
obviously forbidden by relativity.  The second conclusion is that the 
superluminal propagation allowed by the group-velocity definition does 
not give rise to any causal loop paradoxes.  The third conclusion is 
that no fundamental modifications in physics are needed to explain 
these phenomena.  Finally the possibility of interesting applications 
of superluminal signals (in the sense of negative group delays) is an 
open question.

\section*{Acknowledgments}
R.  Y.  Chiao and M.  W.  Mitchell were supported by ONR grant number 
N000149610034.  E.L.  Bolda was supported by the Marsden Fund of the 
Royal Society of New Zealand. It is a pleasure to acknowledge many 
useful conversations with Prof. C. H. Townes, Prof. A. M. Steinberg, 
and Jack Boyce.

\newpage

\end{document}